%% file: paper.tex
\documentclass{article}
\PassOptionsToPackage{numbers,sort&compress}{natbib}
\usepackage[preprint]{neurips_2026}
\usepackage[T1]{fontenc}
\usepackage{courier}
\usepackage[utf8]{inputenc}
\usepackage{graphicx,booktabs,amsmath,amssymb,microtype,xcolor}
\usepackage[colorlinks=true,urlcolor=blue,citecolor=blue,linkcolor=blue]{hyperref}
\input{current-benchmark-values.tex}
\title{Orukeet: Multilingual ASR\\with Frozen Gabor Kernels}
\author{Nathan Roll\textsuperscript{1,2}\quad
Irene Yi\textsuperscript{1,2}\quad
B\"u\c{s}ra Mar\c{s}an\textsuperscript{1,2}\\[3pt]
\bfseries Vianney Grenez\textsuperscript{1}\quad
Gabriel Stein\textsuperscript{4}\quad
Momcilo Mrkaic\textsuperscript{5}\\[3pt]
\bfseries Pavle Padjin\textsuperscript{5}\quad
Vladimir Zeljkovic\textsuperscript{5}\quad
Calbert Graham\textsuperscript{1,3}\\[10pt]
\begin{tabular}{@{}ccccc@{}}
\parbox[c][19pt][c]{0.74in}{\centering\includegraphics[width=0.63in]{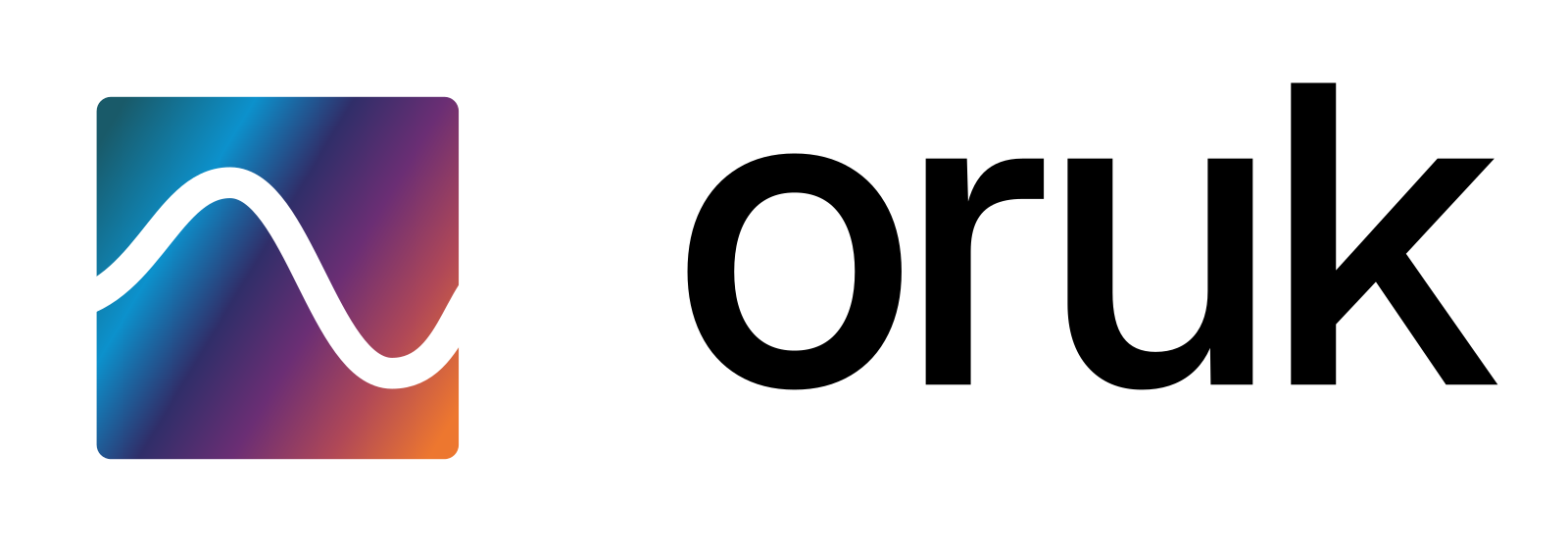}} &
\parbox[c][19pt][c]{1.08in}{\centering\includegraphics[width=1.02in]{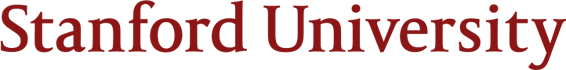}} &
\parbox[c][19pt][c]{1.26in}{\centering\includegraphics[width=1.08in]{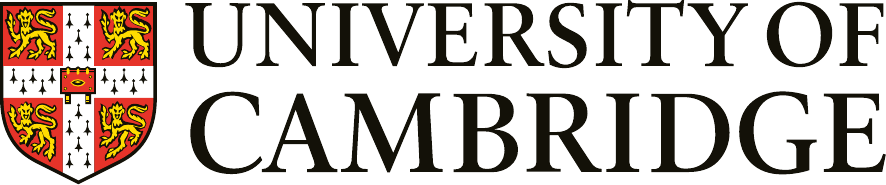}} &
\parbox[c][19pt][c]{0.78in}{\centering\includegraphics[height=17pt]{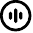}} &
\parbox[c][19pt][c]{0.70in}{\centering\includegraphics[width=0.55in]{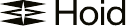}}\\[3pt]
\scriptsize\textsuperscript{1}Oruk AI &
\scriptsize\textsuperscript{2}Stanford University &
\scriptsize\textsuperscript{3}University of Cambridge &
\scriptsize\textsuperscript{4}OpenWhispr &
\scriptsize\textsuperscript{5}Hoid
\end{tabular}}
\hypersetup{pdftitle={Orukeet: Multilingual ASR with Frozen Gabor Kernels},
pdfauthor={Nathan Roll, Irene Yi, B\"u\c{s}ra Mar\c{s}an, Vianney Grenez, Gabriel Stein, Momcilo Mrkaic, Pavle Padjin, Vladimir Zeljkovic, Calbert Graham},
pdfsubject={Orukeet r3 technical report},
pdfkeywords={automatic speech recognition, multilingual speech, Gabor kernels, Parakeet, Orukeet}}
\date{}

\begin{document}
\maketitle
\begin{abstract}
Orukeet replaces half of an adapted Parakeet encoder's temporal filters with
12,288 fitted Gabor kernels, freezes these replacements, and trains the remaining
parameters on multilingual and multi-accent data. Final adaptation and
checkpoint selection use LibriSpeech test-other. Across 20,146 FLEURS recordings
in 25 languages, pooled word error rate (WER) falls from Parakeet's
\FleursPooledParakeetWER\% to Orukeet's \FleursPooledOrukeetWER\%, a
\FleursPooledReduction\% relative reduction. Orukeet has lower WER on
\FleursWins\ of the 25 languages. Orukeet outperforms Parakeet on
\TestedWins\ out of \TestedSplits\ tested splits, including LibriSpeech
test-clean (\LibriCleanOrukeetWER\% vs.\ \LibriCleanParakeetWER\% WER),
test-other (\LibriOtherOrukeetWER\% vs.\ \LibriOtherParakeetWER\%), and
FLEURS English (\FleursEnglishOrukeetWER\% vs.\ \FleursEnglishParakeetWER\%).
All comparisons decode the same audio with matched NeMo settings.
The fitted kernels are stored as ordinary convolution weights, retaining
Parakeet's architecture and inference operators.
\end{abstract}

\section{A fixed structure inside a learned recognizer}
Orukeet fixes selected temporal filters to fitted Gabor functions and trains
the rest of the recognizer. The starting architecture is
NVIDIA's Parakeet TDT 0.6B v3, a 25-language speech recognizer
\citep{parakeet}. Its FastConformer encoder has 24 blocks, each containing
1,024 nine-tap temporal depthwise kernels, and a token-and-duration transducer
(TDT) predicts text \citep{fastconformer,tdt}. We fit a separate Gabor function
to each kernel in a multilingual adaptation of this model, replace the closest
half, and hold the replacements fixed throughout subsequent training.

\begin{figure}[!ht]
\centering\includegraphics[width=\linewidth]{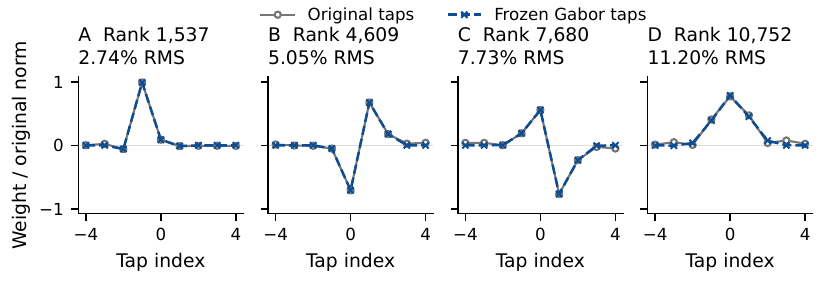}
\caption{Original kernels and fitted Gabor replacements at four predetermined
ranks spanning the selected half. Each pair is divided by the original kernel's
$L_2$ norm; percentages give relative RMS fit error. Markers show the nine
stored taps, joined by straight lines.}
\label{fig:fits}
\end{figure}

The choice is made from the learned filters themselves. A close fit keeps the
initial change small; the remaining weights then accommodate that change.
Figure~\ref{fig:fits} shows the stored taps at four predetermined ranks in the
selected half. We compare the final checkpoint with stock Parakeet across
read speech, accents and domains.

\section{Fitting, freezing and adaptation}
For a nine-tap kernel $w_i$, we fit a Gabor function at
$t\in\{-4,\ldots,4\}$ and rank its normalized squared error:
\begin{equation}
 g_i(t)=A_i\exp\!\left[-\frac{(t-\mu_i)^2}{2\sigma_i^2}\right]
 \cos\!\left(2\pi f_i(t-\mu_i)+\phi_i\right),\qquad
 e_i=\frac{\|w_i-g_i\|_2^2}{\|w_i\|_2^2}.
 \label{eq:gabor}
\end{equation}
For fixed center $\mu$, width $\sigma$ and frequency $f$, linear least squares
solves for the cosine and sine coefficients, giving amplitude $A$ and phase
$\phi$. We evaluate 3,321 initial combinations and refine the best four plus
the best in each frequency quartile in float64, with at most 160 evaluations
per refinement. The search bounds are $\mu\in[-4,4]$,
$\sigma\in[0.25,36]$ and $f\in[10^{-6},0.499999]$ cycles per encoder timestep.
We keep the best evaluated fit and select the 12,288 lowest $e_i$ globally,
with layer and channel as deterministic tie-breaks. A replacement contains
only the fitted function, with no offset or learned residual.

\begin{figure}[!ht]
\centering\includegraphics[width=\linewidth]{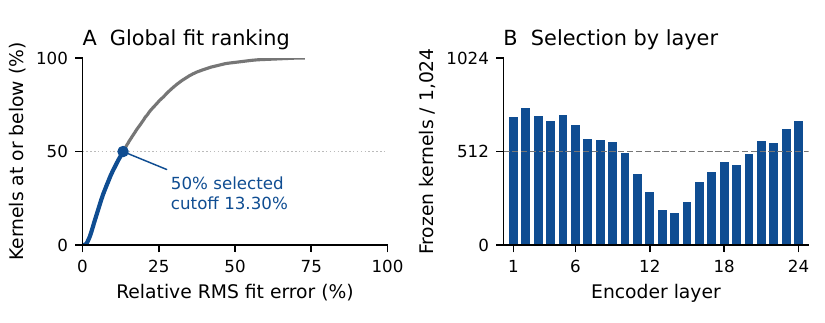}
\caption{Global fit ranking and its allocation across layers. Left: cumulative
relative RMS error for all 24,576 kernels, with the selected half in blue.
Right: selected kernels in every encoder layer; the dashed line marks 512
kernels. Selection uses one global ranking, so each layer need not be half fixed.}
\label{fig:selection}
\end{figure}

This ranking gives 175--748 fixed kernels per layer (Figure~\ref{fig:selection}).
The selected fits have 6.32\% median relative RMS error, a 13.30\% cutoff,
and pooled squared error equal to 0.4244\% of their original weight energy.
The 50\% constraint applies to temporal depthwise kernels. The materialized
model retains 627,008,134 scalar parameters; 110,592 stored taps are fixed and
626,897,542 remain trainable. Analytic audio filters have also been used in
learnable frontends \citep{leaf,sincnet}; here, each function approximates a
learned kernel inside the encoder.

Recovery trains the remaining network with transducer loss, teacher matching
at block and convolution outputs, and token/duration distillation. Subsequent
adaptation applies 4,035 AdamW updates with learning rate decaying from
$10^{-6}$ to $10^{-7}$. The final pass starts from those weights and performs
168 updates, with a 3\% warmup and cosine decay from $5\times10^{-6}$ to
$5\times10^{-7}$. AdamW uses $(0.9,0.98)$, weight decay 0.001 and gradient
clipping at 1.0. Microbatches contain at most 16 utterances and 120 padded
seconds; four microbatches form an update. Training uses BF16 on one A100
40 GB, with dropout, augmentation and dithering disabled.

The final pass makes three passes over the 2,939 LibriSpeech test-other
recordings. Targets preserve the parent's casing and punctuation while
correcting reference words; all targets match the reference under the pinned
English normalizer. Test-other also supplies the final checkpoint-selection
comparison. All 651 remaining parameter tensors change. An independent audit
of the exported checkpoint verifies that the 12,288 fitted kernels are
byte-identical to the original fitted functions; tokenizer assets, signal
processing buffers and batch-normalization statistics also remain unchanged.

\section{Recognition across languages, accents and domains}
\paragraph{Matched comparison.}
We restore stock Parakeet and the final Orukeet checkpoint independently and
decode every recording with NeMo greedy-batch TDT, a ten-symbol limit, FP32
weights and BF16 CUDA autocast. Matrix-multiply TF32 is disabled. Both models
receive identical mono 16 kHz audio, in the same duration-sorted batches;
empty transcripts remain in the scores. We compare the resulting Orukeet
checkpoint with pretrained Parakeet on the splits used throughout model
development.

English uses a pinned text normalizer with spelling, number and compound
maps. Other languages retain diacritics and use language-specific number
normalization and compound-boundary alignment. Word errors count substitutions,
deletions and insertions. For a set of recordings $\mathcal D$, pooled WER is
\begin{equation}
 \operatorname{WER}_{\mathrm{pool}}(\mathcal D)
 =100\,\frac{\sum_{u\in\mathcal D}(S_u+D_u+I_u)}
                 {\sum_{u\in\mathcal D}N_u},
 \label{eq:pooled}
\end{equation}
where $N_u$ is the normalized reference word count. Compound alignment can
change this denominator separately for each model. A language macro instead
weights each language's WER equally. Integer counts and independent
full-partition rescoring reproduce every reported value.

\paragraph{Read speech in 25 languages.}
The complete comparison contains 25,705 recordings: both LibriSpeech test
partitions \citep{librispeech} and FLEURS test speech in all 25 supported
languages \citep{fleurs}. Table~\ref{tab:current-standard} reports every partition.
FLEURS pooled WER is \FleursPooledParakeetWER\% for Parakeet and
\FleursPooledOrukeetWER\% for Orukeet; the corresponding language macros are
\FleursMacroParakeetWER\% and \FleursMacroOrukeetWER\%.
Orukeet improves \StandardWins\ of the 27 partitions, including
\FleursWins\ of 25 FLEURS languages.

\input{current-standard-table.tex}

\paragraph{Accents and domains.}
We also evaluate both models on a fixed sample of 12,006 recordings across
47 partitions and 25 languages. It includes EuroSpeech, GigaSpeechBench,
Monsoon, Golos, NST, VoxPopuli and Lesbos: 256 recordings per partition and all
230 Lesbos recordings. The preceding 4,035-update adaptation includes 6,118
of these recordings. Greek and Italian EuroSpeech use the
audited human transcript spans. Both checkpoints are decoded afresh and scored
with the same pinned protocol as the read-speech comparison.
Pooled WER is \DomainPooledParakeetWER\% versus \DomainPooledOrukeetWER\%;
for the 5,120 English recordings, it is \DomainEnglishParakeetWER\% versus
\DomainEnglishOrukeetWER\%. Orukeet improves \DomainWins\ of 47 partitions,
including \DomainEnglishWins\ of 20 English partitions. Table~\ref{tab:current-domains}
gives every score. We pool read speech and accent/domain recordings separately.

\section{Checkpoint and reproduction}
All recognition results in this report refer to the NeMo checkpoint with
SHA-256 prefix \texttt{031c8ddab484}. The file stores the configuration,
tokenizer and materialized convolution weights. The fitting code retains each
kernel's analytic parameters; training uses a fixed parametrization to prevent
updates to the selected rows. Inference uses ordinary depthwise convolution,
with the same tensor shapes and operator counts as Parakeet.

The \href{https://github.com/Oruk-AI/orukeet}{Orukeet repository} and
\href{https://huggingface.co/oruk/orukeet/resolve/555136b50265a132d4cea0d35560c26fc4f657ab/orukeet-v0.1.0.nemo}{release checkpoint}
contain the source weights, training recipes, fitted functions, export audits
and reproducible evaluation records. The metric bundle retains unrounded
scores, per-record edit counts, manifest identities and checkpoint hashes.
Code is MIT; weights and fits are CC BY-SA 4.0; metric records are CC BY 4.0.
NVIDIA's foundation attribution is retained.

\par\begin{minipage}{\linewidth}
\small
\bibliographystyle{plainnat}
\bibliography{references}
\end{minipage}
\input{current-domains-table.tex}
\end{document}

%% file: current-benchmark-values.tex
\newcommand{\LibriCleanParakeetWER}{1.53}
\newcommand{\LibriCleanOrukeetWER}{1.46}
\newcommand{\LibriOtherParakeetWER}{3.14}
\newcommand{\LibriOtherOrukeetWER}{2.86}
\newcommand{\FleursEnglishParakeetWER}{4.28}
\newcommand{\FleursEnglishOrukeetWER}{3.82}
\newcommand{\FleursPooledParakeetWER}{11.01}
\newcommand{\FleursPooledOrukeetWER}{9.85}
\newcommand{\DomainPooledParakeetWER}{16.72}
\newcommand{\DomainPooledOrukeetWER}{15.25}
\newcommand{\DomainEnglishParakeetWER}{9.51}
\newcommand{\DomainEnglishOrukeetWER}{8.84}
\newcommand{\FleursMacroParakeetWER}{11.07}
\newcommand{\FleursMacroOrukeetWER}{9.96}
\newcommand{\FleursWins}{23}
\newcommand{\StandardWins}{25}
\newcommand{\DomainWins}{36}
\newcommand{\DomainEnglishWins}{20}
\newcommand{\TestedWins}{61}
\newcommand{\TestedSplits}{74}
\newcommand{\FleursPooledReduction}{10.6}

%% file: current-standard-table.tex
\begin{table}[!ht]
\centering
\caption{Complete LibriSpeech and FLEURS test partitions. WER and CER are percentages; bold identifies lower WER. The pooled FLEURS row sums errors and reference words over all 25 languages, including English. The macro row weights languages equally.}
\label{tab:current-standard}
\small
\setlength{\tabcolsep}{5pt}
\renewcommand{\arraystretch}{1.02}
\begin{tabular}{lrrrrr}
\toprule
Benchmark & Clips & \multicolumn{2}{c}{Parakeet} & \multicolumn{2}{c}{Orukeet}\\
 & & WER & CER & WER & CER\\
\midrule
LibriSpeech test-clean & 2,620 & 1.53 & 0.59 & \textbf{1.46} & 0.56\\
LibriSpeech test-other & 2,939 & 3.14 & 1.32 & \textbf{2.86} & 1.19\\
\midrule
FLEURS Bulgarian & 658 & 11.92 & 3.84 & \textbf{10.37} & 3.34\\
FLEURS Croatian & 914 & 11.29 & 3.53 & \textbf{10.20} & 3.67\\
FLEURS Czech & 723 & 11.12 & 3.21 & \textbf{8.97} & 2.67\\
FLEURS Danish & 930 & 17.19 & 6.31 & \textbf{14.88} & 5.31\\
FLEURS Dutch & 364 & 6.40 & 2.28 & \textbf{5.60} & 1.93\\
FLEURS English & 647 & 4.28 & 2.00 & \textbf{3.82} & 1.77\\
FLEURS Estonian & 893 & 13.32 & 3.86 & \textbf{10.44} & 3.39\\
FLEURS Finnish & 918 & 11.14 & 2.59 & \textbf{9.35} & 2.16\\
FLEURS French & 676 & \textbf{4.69} & 1.68 & 5.01 & 1.70\\
FLEURS German & 862 & 4.21 & 1.41 & \textbf{3.92} & 1.52\\
FLEURS Greek & 650 & \textbf{21.07} & 9.01 & 30.81 & 9.18\\
FLEURS Hungarian & 905 & 13.60 & 4.20 & \textbf{10.68} & 2.97\\
FLEURS Italian & 865 & 2.43 & 0.79 & \textbf{2.09} & 0.76\\
FLEURS Latvian & 851 & 21.78 & 5.43 & \textbf{17.41} & 4.21\\
FLEURS Lithuanian & 986 & 20.95 & 5.56 & \textbf{16.55} & 4.27\\
FLEURS Maltese & 926 & 19.22 & 6.19 & \textbf{15.60} & 5.08\\
FLEURS Polish & 758 & 6.81 & 2.09 & \textbf{6.11} & 1.95\\
FLEURS Portuguese & 919 & 4.49 & 1.98 & \textbf{3.73} & 1.63\\
FLEURS Romanian & 883 & 11.44 & 3.86 & \textbf{9.34} & 3.07\\
FLEURS Russian & 775 & 4.89 & 1.49 & \textbf{4.72} & 1.48\\
FLEURS Slovak & 792 & 9.21 & 2.91 & \textbf{7.75} & 2.41\\
FLEURS Slovenian & 834 & 22.62 & 7.70 & \textbf{22.11} & 8.28\\
FLEURS Spanish & 908 & 3.22 & 1.28 & \textbf{2.75} & 1.04\\
FLEURS Swedish & 759 & 13.38 & 4.26 & \textbf{11.36} & 3.45\\
FLEURS Ukrainian & 750 & 6.00 & 1.74 & \textbf{5.39} & 1.60\\
\midrule
FLEURS pooled & 20,146 & 11.01 & 3.57 & \textbf{9.85} & 3.13\\
FLEURS language macro & 20,146 & 11.07 & 3.57 & \textbf{9.96} & 3.15\\
\bottomrule
\end{tabular}
\end{table}

%% file: current-domains-table.tex
\begin{table}[!p]
\centering
\caption{WER (\%) on the fixed accent and domain sample. Each partition contains 256 recordings, except Lesbos (230). GSB denotes GigaSpeechBench; two-letter suffixes identify languages. Bold identifies lower WER. Both models use the same decoding and scoring protocol as Table~\ref{tab:current-standard}.}
\label{tab:current-domains}
\small
\begin{minipage}[t]{0.49\linewidth}
\vspace{0pt}
\centering
\setlength{\tabcolsep}{3pt}
\renewcommand{\arraystretch}{1.10}
\begin{tabular}{lrr}
\toprule
Partition & Parakeet & Orukeet\\
\midrule
EuroSpeech BG & 14.22 & \textbf{13.04}\\
EuroSpeech DE & 13.40 & \textbf{11.14}\\
EuroSpeech EL & \textbf{25.83} & 26.35\\
EuroSpeech EN & 24.40 & \textbf{23.77}\\
EuroSpeech ET & 34.67 & \textbf{25.33}\\
EuroSpeech FI & 16.61 & \textbf{15.20}\\
EuroSpeech FR & 19.42 & \textbf{14.28}\\
EuroSpeech HR & 12.93 & \textbf{12.56}\\
EuroSpeech IT & \textbf{10.95} & 12.32\\
EuroSpeech LT & 38.44 & \textbf{33.10}\\
EuroSpeech LV & 57.18 & \textbf{42.14}\\
EuroSpeech MT & 36.83 & \textbf{36.15}\\
EuroSpeech PT & \textbf{23.08} & 23.81\\
EuroSpeech SK & 17.29 & \textbf{14.91}\\
EuroSpeech SL & \textbf{48.43} & 50.23\\
EuroSpeech UK & \textbf{13.65} & 14.25\\
GSB AI & 8.71 & \textbf{7.98}\\
GSB Chinese accent & 14.49 & \textbf{13.56}\\
GSB Filipino accent & 13.30 & \textbf{12.79}\\
GSB Indian accent & 6.50 & \textbf{5.59}\\
GSB Japanese accent & 19.15 & \textbf{17.78}\\
GSB Scottish accent & 22.08 & \textbf{20.35}\\
GSB Singaporean accent & 13.89 & \textbf{12.86}\\
GSB agriculture & 6.20 & \textbf{5.84}\\
\bottomrule
\end{tabular}
\end{minipage}\hfill%
\begin{minipage}[t]{0.49\linewidth}
\vspace{0pt}
\centering
\setlength{\tabcolsep}{3pt}
\renewcommand{\arraystretch}{1.10}
\begin{tabular}{lrr}
\toprule
Partition & Parakeet & Orukeet\\
\midrule
GSB arts & 5.47 & \textbf{4.87}\\
GSB biology & 3.67 & \textbf{3.31}\\
GSB economics & 7.05 & \textbf{6.57}\\
GSB engineering & 4.06 & \textbf{3.50}\\
GSB entertainment & 10.40 & \textbf{8.87}\\
GSB finance & 5.81 & \textbf{5.11}\\
GSB humanities & 7.98 & \textbf{7.47}\\
GSB law & 9.75 & \textbf{9.04}\\
GSB medicine & 3.49 & \textbf{3.18}\\
GSB military & 3.43 & \textbf{3.06}\\
Golos crowd RU & \textbf{2.84} & 2.92\\
Golos far-field RU & \textbf{7.98} & 9.10\\
Lesbos Greek & 94.78 & \textbf{93.55}\\
Monsoon India & 4.12 & \textbf{3.78}\\
NST Danish & 26.49 & \textbf{11.59}\\
NST Swedish & 16.57 & \textbf{12.36}\\
VoxPopuli CS & \textbf{7.32} & 7.39\\
VoxPopuli ES & \textbf{6.07} & 6.20\\
VoxPopuli HU & 12.00 & \textbf{11.05}\\
VoxPopuli IT & \textbf{11.37} & 11.82\\
VoxPopuli NL & \textbf{9.50} & 9.56\\
VoxPopuli PL & 6.48 & \textbf{6.24}\\
VoxPopuli RO & 11.48 & \textbf{11.20}\\
\bottomrule
\end{tabular}
\end{minipage}
\par\vspace{12pt}
\begin{tabular}{lrrr}
\toprule
Pooled comparison & Clips & Parakeet & Orukeet\\
\midrule
All 47 partitions & 12,006 & 16.72 & \textbf{15.25}\\
All 20 English partitions & 5,120 & 9.51 & \textbf{8.84}\\
\bottomrule
\end{tabular}
\end{table}